\documentstyle[psfig,twocolumn,prb,aps]{revtex}
\begin{document}
\title{
\draft
Solitonic approach to the dimerization problem in 
correlated one-dimensional systems }

\author{ 
Ji\v{r}i M\'alek \cite{mal}, Stefan-Ludwig Drechsler \cite{dre}, and Gernot
Paasch }
 \address{ Institut f\"ur Festk\"orper- und Werkstofforschung Dresden e.V.,
 Postfach 270016, D-01171 Dresden, Germany }

\author{Karen Hallberg} 
\address{ Centro At\'omico Bariloche,
8400 San Carlos de Bariloche, Argentina and \\
Max-Planck-Institut f\"ur Physik komplexer Systeme, Bayreuther 
 Str.\ 40, D-01187 Dresden, Germany}
\date{\today}

\maketitle
\widetext
\centerline{\parbox[c]{14cm}{\small Using exact diagonalizations
we consider  self-consistently the lattice distortions in odd Peierls-Hubbard
and spin-Peierls periodic rings  in the adiabatic harmonic approximation.
From the tails of the inherent spin soliton the dimerization $d_{\infty}$ of
regular even rings is found by extrapolations to infinite ring lengths.
Considering a wide region of
electron-electron onsite interaction values 
$ U>0 $  compared with the band width                        
$4t_0$ at intermediately strong electron-phonon  interaction $g$,
known relationships obtained  by other methods are reproduced and/or refined
within one unified approach: such as the 
maximum of  $d_{\infty}$ at $U$$\simeq$3$t_0$ for
$g$$\simeq$0.5 and its shift to zero for $g$$\to$$g_c$$\approx$0.7.
The hyperbolic tangent shape of the spin soliton is retained for
any $U$ and $g\stackrel{<}{\sim}$0.6.
In the spin-Peierls limit the  $d_{\infty}$ are found to be in agreement
with results of DMRG computations.}}
\hspace{2cm}
\pacs{63.20.Kr, 71.20.Rv, 71.27+a, 71.45.Lr}

\vspace{0.1cm}

\normalsize
\narrowtext
There is a longstanding debate on the 
interplay of the electron-electron ({\it el-el}) 
interaction and the electron-phonon ({\it el-ph}) interaction 
in conducting polymers like {\it trans}-(CH)$_x$.
Among a large amount of papers we refer here only to 
Refs.\  1-6 
devoted   to the discussion of the 
origin of the
observed dimerization $d$ 
in the framework of   the (extended) 
Peierls-Hubbard model (PHM). The PHM is  
regarded as the minimal microscopic model
for conducting polymers.
Special features of the interplay of on-site correlation $U$ 
and off-diagonal {\it el-ph} interaction have been pointed out first 
in Ref.\ 1. Employing the Gutzwiller approximation (GA),  
it was shown 
 that for weak and intermediate 
{\it el-ph} interaction strength, the dimerization  $d$
 passes through  a maximum near $U\approx$4$t_0$ when $U$ is increased
 after which
 it is 
 suddenly suppressed.
The geminal
approach \cite{kuprievich94} (GEA) shows a smooth decrease of $d$
with increasing $U$ 
in the opposite strongly correlated limit. According to 
extrapolations based on exact diagonalizations (ED)
\cite{dixit84}$^,$\cite{hayden88}$^,$\cite{waas90}  the 
enhancement of $d$ due to $U$
predicted by the GA near the maximum is overestimated. 
In Refs.\ 2,3 the infinite chain limit $d_{\infty}$
calculated using
the PHM     
was    extrapolated from 
above (below) by $2n$-membered open chains ($4n$+2-
membered rings).
In the highly correlated limit of
the 1/2-filled band case  
the PHM can be mapped
onto the antiferromagnetic spin-1/2 Heisenberg model
(AFHM) and at low temperatures  a spin-Peierls phase is expected.
Such dimerized phases have been observed for
CuGeO$_3$ (having CuO$_2$ chains)
and $\alpha ^{\prime} $-NaV$_2$O$_5$. The 
CuO$_3$ chains in Sr(Ca)$_2$CuO$_3$ are at the threshold to the 
AFHM-limit\cite{rosner97}.  In many cases of practical
interest the
actual $d$ is very small and no self-consistent $d\neq$0 is found in
feasible short {\it even} rings. 

In the present paper we show that studies of {\it odd} rings
of comparable  lengths yield reliable estimates of lattice

.
\vspace{4.57cm}

\noindent
distortions in infinite rings at arbitrary strength of {\it el-el} correlations
and reasonable strength of the {\it el-ph} interaction.
We exploit the generic property of neutral odd periodic rings that their ground
state is given by a spin soliton. The lattice distortions for such 
a ring are shown schematically in Fig.\ 1.
The bond in front of the soliton center, i.e. between sites $N$ and 1,
is a long bond for which $a_{N1}$-$a_0$=$2u_0\equiv$$d$$>0$ holds ( 
$a_{N1}$ ($a_0$) is  the  bond-length in the distorted (equidistant) state).
For large $N$ the region far from the soliton center tends to the regularly
dimerized state. Hence, varying the model parameters, insight into the behavior
of $d_{\infty}$ may be gained already at finite $N$ from the study of $d$.
For short rings ($N=3,5,.$), $d(N)$ exceeds significantly
$d_{\infty}$.  This is  the consequence of a strong first-order Jahn-Teller
effect \cite{soos85}. In Refs.\ 8,9 the total energy $E_{\rm tot}$ of odd AFHM
rings has  been studied for {\it fixed} geometries adopting rigid
sharp soliton shapes. In our method {\it all} bonds are optimized to yield the
minimum of the total energy $E_{\rm tot}$. The inherent soliton exhibits a
{\it smooth} shape.  We illustrate our method
considering  the PHM and  the AFHM, where comparison with other reliable
approaches is possible.  In particular, the  density matrix renormalization
group is used to check the extrapolated $d_{\infty}$.

For the electronic part $H_{\rm el}$ of the total Hamiltonian 
$H=H_{\rm el}+H_{\rm lat}$ we adopt the  one-band extended PHM 
\begin{eqnarray}
H_{\rm el}=& \sum_{i,s} t_{i,i+1}\left( c^{\dag}_{i,s}c_{i+1,s}+{\rm H.c}.
\right)
+ \nonumber \\ 
\label{model}\\
& +\sum_{i}\left(Un_{i,\uparrow}n_{i,\downarrow} + 
V_{i,i+1}n_in_{i+1}\right),\nonumber \\
\nonumber 
\end{eqnarray}
\noindent  
at half filling, 
where $c^{\dag}_{i,s}$ creates an electron with spin $s$=$\pm1/2$ at
site $i$, $n_{i,s}$=$c^{\dag}_{is}c_{is}$ is the number
operator, and $n_i$=$\sum_sn_{is}$.
We linearize the bond-length dependent transfer integral $t$ and the 
intersite {\it el-el}
interaction $V$
\begin{equation}
t_{i,i+1}=-\left(t_0-\gamma v_i\right), \quad V_{i,i+1}=V-\eta v_i,
\label{linearized}
\end{equation}
where $v_i$=$u_{i+1}$-$u_i$, 
$u_i$ is the displacement of the $i^{\mbox{\small th}}$ site relative to
the  undistorted state.
In the adiabatic and harmonic approximation 
the lattice part $H_{\rm lat}$ reads as
\begin{equation}
H_{\rm lat}=(K/2)\sum_iv_{i}^2,
\label{elastic}
\end{equation}
where $K$  is the spring constant.
Via the Hellmann-Feynman theorem we  obtain $N$ self-consistent Eqs.\  
\begin{equation} 
Kv_i= \Lambda /N - 2\gamma P_{i,i+1} + \eta D_{i,i+1} ,
\label{self}
\end{equation}
where  $\Lambda$=$\sum_i\left(2\gamma P_{i,i+1} - \eta D_{i,i+1} \right)$ 
expresses the fixed length constraint $\sum_i v_i =0$,
$P_{i,i+1}$ being the bond order, and $D_{i,i+1}$ denotes the     
density-density correlators in the ground state  $\mid G\rangle$  
\begin{equation}
P_{i,i+1}=\frac{1}{2}\sum_s \langle G\mid c^{\dagger}_{i,s}c_{i+1,s}+
{\rm H.c}. \mid
G\rangle,
\end{equation}
\begin{equation}
D_{i,i+1}=\langle G\mid n_{i}n_{i+1} \mid G\rangle.
\end{equation}

The strength of the {\it el-ph} interaction can be measured 
by the parameter $g$ introduced as \cite{waas90}
\begin{equation}
g=\gamma /\sqrt{Kt_0} < 1.
\label{elph}
\end{equation}
 Note that sometimes \cite{baeriswyl84} the related quantities
 defined as 
\begin{equation}
 \lambda_{\rm el-ph}=2g^2/\pi, \quad \mbox{or} \quad
 \lambda_{\rm SSH}=2\lambda_{\rm el-ph}.
\label{lambda}
\end{equation}
 are used.
 Values of $g\sim$ 0.4 to 0.5 typical for conducting polymers are regarded as
 weak to intermediate coupling constants. Following Refs.\ 5,6 we shall use
 hereafter the  dimensionless dimerization defined as 
\begin{equation}
 d=2u_0\sqrt{K/t_0} < 1.
\label{dimerization}
\end{equation}
 The parameter $\delta $ frequently used to  describe the modulation of the
transfer integral $t_{i,i+1}=t_0(1+(-1)^{i+1}\delta )$ in the dimerized
state \cite{baeriswyl84}$^-$ \cite{hayden88} is related to $d$ by $\delta =gd$.
 In the limit $U/t_0$$\gg 1$ the low-energy physics of 
 the 1/2-filled PHM ring (Eq.\ (\ref{model})) can be described by the AFHM
\begin{equation}
H_{\rm sp}=\sum_{i} J_{i,i+1}\vec{S}_i\vec{S}_{i+1}
,\ \mbox{with} \ 
  J_{i,i+1}\approx\frac{4t_{i,i+1}^2}{U-V_{i,i+1}}
\label{genheis}
\end{equation}
where the exchange integral is given to 2$^{nd}$ order perturbation theory 
 in $t_0/U$. Using Eq.~(2) we find 
\begin{equation}
J_{i,i+1}=J_0-\gamma_{\rm sp} v_i=J_0(1+(-1)^{i+1}\delta_{\rm sp}),
\end{equation}
where $\gamma_{\rm sp}$=2$\gamma$$J_0/t_0$ and $\delta_{\rm sp}$=2$gd$
characterizes the regular spin-Peierls state for $\eta$=0.
We shall use the  AFHM also  out of the limit $U$$\gg$$t_0$ for the case
$V$=0 adopting an effective exchange  integral $J$=$J(U/t_0)$ given by the
relation $J_{ij}$=(2/$\pi$)$v_{\rm sp}t_{ij}$ and the spin
velocity $v_{\rm sp}$ taken from the Bethe-Ansatz solution for the equidistant
infinite Hubbard ring \cite{ovchinnikov69} (hereafter $a_0$=1, $\hbar$=1)
\begin{equation}
J_{i,i+1}=
\frac{4}{\pi}\frac{t_{i,i+1}I_1(z_{i,i+1})}{I_0( z_{i,i+1})}, \quad
z_{i,i+1}=\frac{2}{\pi U} t_{i,i+1},
\label{effex}
\end{equation}
where $I_n$, ($n=0,1$) are modified Bessel functions.
 
Using the  L\'anczos-method, the Hamiltonians (Eqs.\ (\ref{model},
\ref{genheis})) have been diagonalized exactly for finite rings with periodic
boundary conditions starting with a given set $\{ v_{n,(0)} \}$.
The AFHM has been treated using the spinless fermion technique \cite{fradkin91}
resulting in analogous self-consistent equations as for the PHM. Then the
corresponding ``ground-state'' eigenvector $\mid G\rangle$ has been used to
calculate the next set of lattice order parameters $\{v_{n,(1)}\}$ using Eq.\
(\ref{self}). The iteration was continued until the maximal deviations of
$E_{\rm tot}$ and all $v_{n,(j)}$ between two iteration steps $j$ and $j+1$
became smaller than the required accuracy of $10^{-7}$ to  $10^{-8}$ 
(see Refs.\ 3,12). Here rings composed of up to  $N$=13 sites (PHM) and $N$=23
(AFHM) have been  studied. The ED computer limitations to rings where finite
size effects are still important can be circumvented at least for {\it even}
membered regularly dimerized AFHM rings and reasonable {\it el-ph} ({\it sp-ph}) interaction strength applying the density matrix renormalization group
technique \cite{white92} (DMRG) with typical discarded errors of the order
10$^{-6}$.

At first we consider the one-particle Su-Schrieffer-Heeger model  
(SSH) (PHM: $U$,$V$=0) where very long odd 
rings can be treated numerically. The infinite even ring problem is reduced
to a transcendental equation for $d_{\infty}$:
\begin{equation}
\frac{1}{\lambda_{\rm SSH}}=\frac{K(k)-E(k)}{k^2},
\quad \mbox{with} \quad k=\sqrt{1-(gd_{\infty})^2},
\end{equation}
where $K$ and $E$ denote complete elliptical integrals.
The results for up to $N=601$ sites are shown in Fig.\ 2.
Starting from small $N$, $d$ passes    at first by  
 a minimum at $N$ $\simeq$ 2$\xi$, the width of a soliton in an infinite ring.
For weak {\it el-ph} coupling $\xi \gg 1$,
$d_{\rm min}/d_{\infty} \approx 0.86 $ at $2\xi /N \approx$ 0.92. 
Below that minimum all curves, for which $N > 2\xi $ holds, approach a
nearly universal curve, pass a very small maximum near $6\xi$, and tend finally
to $d_{\infty}$ from above\cite{remark8}. Since for  correlated problems  only
 relatively short rings can be treated exactly, a strong nonmonotonous 
 behavior  could cause problems in  extrapolating to $d_{\infty}$.
Fortunately, our calculations indicate  that at least in the
AFHM-limit the depth of the first minimum at finite $N$ is strongly suppressed.

Let us now consider how the on-site interaction $U$ affects $d$.
As shown in Fig.\ 3 starting from $U=0$, $d$
increases with $U$ and has a maximum at $U_{\rm max} \sim 3t_0$, after which 
$d$ starts to decrease  smoothly. Both behaviors are similar to the GA and GEA
predictions, respectively. Quantitatively, however, we obtain
$U_{\rm max}(g)$$\approx$3.12$t_0$ for $g\sim$0.5. For 0.5$<g<$0.6,
$U_{\rm max}$ starts to decrease. Finally, when $g\rightarrow$
$g_c\approx$0.7,
$U_{\rm max} \rightarrow 0$. For comparison we note that the GA's results are
 $U_{\rm max} \approx 4t_0$ for $g < g_{\rm c}$=0.76. The ED-results of
   Ref.\ 5 
 yield $g_{\rm c}$=0.75$\pm$ 0.04 slightly above our result.
Above  $g_{\rm c}$, there is no enhancement of $d$ due to $U$.
Turning to larger rings, we first adopt an $1/N$ extrapolation (dashed curve) 
and arrive at a rough lower bound being crudest for small $U$.  An improved 
bound is achieved connecting the exact $U$=0 point with the $1/N$ extrapolation
upshifted to the AFHM limit (see below). 
 To avoid an artificial minimum, we omit the $U$=$t_0$ point.
Thus for $g$=0.5 we get 0.174$<$$d_{\infty,{\rm max}}$$\stackrel{<}
{\sim}$$d_{N=13}$
$\approx$0.2. Instead the
 GA\cite{kuprievich94} gives $d_{\rm max}$$\approx$0.31.
For $U\gg t_0$, $g\leq$0.5 the $d_{\rm PHM}$$\to$$d_{\rm AFHM}$ from
above. To compare our $d$ with the continuum model result
of Inagaki et {\it al.}  \cite{inagaki83},
we rewrite their dimerization  as 
\begin{equation}
d = \frac{g^2}{\pi\sqrt{1+\kappa}}\left(\frac{\partial J_0}
{\partial t_0}\right)^2
\left(\frac{t_{0}}{J_0}\right)^{1/2} \rightarrow
cg^2\left(\frac{4t_0}{U-V}\right)^{3/2},
\label{dimu}
\end{equation}
where $c$=$(4/\pi)\sqrt{2/3}$$\approx$1.04 for $\kappa=0.5$ 
(see Eq.\ (10) and Refs.\ 16 -18).
Applying Eq.\ (\ref{dimu}) to intermediately correlated cases,
we adopt the effective exchange integral $J$ defined by Eq.\
(\ref{effex}) and arrive at an {\it analytic} expression (the dashed-dotted
curve in Fig.\ 3) \cite{remark4}. Surprisingly it exhibits similar shape and
magnitude as the weak coupling ($g \leq 0.5$) ED-curves. In particular its
$U_{\rm max}/t_0$=3.21 is close to 3.12 mentioned above. This suggests that even in
the case of conducting polymers, being clearly outside the usual AFHM-regime,
the dimerization is mainly governed by the (always present) 
spin degrees of freedom and to less extent by the charge degrees of freedom. 
In the usual case $U>$2$V$, the $V$ and its derivative $\eta$ (see Eq.\
(\ref{linearized})) enhance $d$. For $U\gg$$t_0$,$V$ and $\eta$=0 one can
replace $U$$\rightarrow$ $U$--$V$.

For long rings $N > 2\xi$, $d_N$-values close to $d_{\infty}$ can be expected.
Then $d=d(N)$ might be approximated by 
\begin{equation}
d(N)=d_{\infty} +\sum_{l=1}^{l_{max}}\frac{A_l}{N^l}\exp\left(\frac{-N}{2 \xi}\right) \quad
+....\quad . 
\label{dsize}
\end{equation}
Note that in contrast with the general PHM case, $d(N)$ for 4$n$ and 4$n$+2
AFHM-rings can be described by {\it one} smooth  curve \cite{soos85}.
The $d_{\rm even} \rightarrow$ $d_{\infty}$  from below, just opposite to  odd
rings. To be specific, we consider one typical example. We estimate for the 
upper curve shown in
Fig.\ 4, $d_{{\rm odd},\infty}=0.0765$, $l_{max}=1, A_{1,{\rm odd}}$$\approx$0.5 and
$\xi_{\rm odd}$=5.26. The even ring curve tends from below to a slightly
larger value $d_{{\rm even},\infty}$=0.078 and $A_{1,{\rm even}}$=--11.8.
The exponent $\xi_{\rm even}$=1.85  differs significantly from $\xi_{\rm odd}$.
From the soliton shape $(-1)^nv_n$$\approx$$d_{N}
\tanh (n/\xi_N)$, we deduced at $N$=23, $\xi_N$$\approx$5, $d_N$$\approx$0.08,
whereas the continuum model \cite{nakano81} yields
$\xi$=$3\pi t_0/(16J_0g^2)$=4.91.
The fit of the even curve can be somewhat 
improved adopting $l_{max}=4$. Then with
$A_1=A_3=0$ and $A_2=-3.5, A_4=-2040$  one arrives at  the same $\xi=5.26$ as
in the odd case for $l_{max}=1$.
Taking the DMRG-values for $N$=60, we conclude that 
the accuracy of the solitonic estimate of $d_{\infty}$ is $\sim  
2$ to $3$ \%. The continuum theory\cite{inagaki83}, (Eq.\ (\ref{dimu}))
 predicts, for $\kappa$=0.5, $d_{\infty}$=0.07203, a value 
slightly below our discrete results. According to our numerical finding
we would recommend to use $\kappa \approx$0.279. 
Fitting alternatively the curvature of $d_{\rm odd}$  at large $N$ by a
parabola, one arrives  at an extrapolated very shallow
minimum at {\it finite} ring length ($N_{\rm min}$$\approx$29 sites for the
present parameters).  Then the slightly smaller $d_{\infty,{\rm odd}}$
compared with $d_{\infty ,{\rm even}}$ might be viewed as a hint for
a tiny minimum at {\it finite} $N$ generic for $d_{odd}$ being the
deepest in the SHH case (see Fig.\ 2). Raising $g$, 
the soliton becomes narrower. Thus at large $g$ any minimum 
should be accessible by the ED. With increasing $g$ 
the $d_{\rm odd}(1/N)$ curves become  flatter. Small minima 
were detected for $g$=0.9, 0.85 at $N$=$N_{\rm min}$=17, 21, respectively.
Anyhow,
the $1/N$-extrapolation of $d_{\rm odd}$ ($d_{\rm even}$) from accessible
$N$$\leq$$N_{\rm min}$ yields a lower (upper) bound of $d_{\infty}$.

To summarize, a novel approach to the dimerization problem of
correlated 1D-models has been presented.
It is based on exact diagonalizations of odd ring Hamiltonians combined
with a self-consistent treatment of the 
classical lattice degrees of freedom. Known dependences of the
bond alternation on the {\it el-el} and {\it el-ph} coupling strengths
obtained by other   approximations  valid in different
parameter regimes have been reproduced and refined within one
 unified method. The  $1/N$-extrapolation to the infinite
 rings gives  a new lower bound for any correlation 
strength. With the aid of the Bethe-Ansatz solution for the spin velocity,
even in the intermediate coupling regime 
a sizeable part of the dimerization can be described by an effective
spin-Hamiltonian gaining   thus new insights in the dimerization 
mechanism of conducting polymers. The DMRG is found out as a  valuable
supplementary tool to our solitonic method.

The Deutsche Forschungsgemeinschaft and the Max-Planck-Institut f\"ur 
Physik komplexer Systeme Dresden are acknowledged
for financial support (S.-L.\ D.\ and J.\ M.) under projects  Dre-269 and
Pa-470. K.H.\ is supported by CONICET.

\newpage
\begin{figure}
{FIG.\ 1\ \small 
Schematical view of the lattice distortions in
1/2-filled odd rings. (Un)distorted sites are
denoted by $\circ$($\bullet$). 
\label{fig1}}
\end{figure}
\begin{figure}
\figure{FIG.\ 2\ \small Reduced dimerization in odd SSH rings  $vs.$ reciprocal
ring length $1/N$ for various {\it el-ph} interactions $g$. The ring length $N$
is given in units of the soliton width $2\xi$=2/$(gd_{\infty})$.
\label{fig2}}
\end{figure}
\begin{figure}
\figure{FIG.\ 3 \small Dimerization in the Peierls-Hubbard model
{\it vs.} on-site energy $U$  for  the {\it el-ph} coupling $g$=0.5. 
In deriving the improved lower bound, Eqs.\ (13-15) have been used.
\label{fig3}}
\end{figure}
\begin{figure}
\figure{FIG.\ 4 \small 
Size dependence of the dimerization $d$ for even ($\circ$) and odd ($\bullet$)
periodic spin-Peierls rings. Even rings are treated by ED until $N$=22. The $d$
for $N$=28 to 60 are obtained by the DMRG. The parameter set used $\gamma_{\rm
sp}$=0.4, $J$=1/3, $K$=1 corresponds to $U$=13,$V$=$t_0$=1,$\eta$=0, and
$g$=0.6 for the PHM. The full and the dashed curves are the fits by Eq.\
(\ref{dsize}).
\label{fig4}}
\end{figure}

\begin{references}
\bibitem[14]{remark8}
Strictly speaking, there are two slightly different smooth
curves for 4$n\pm 1$ membered rings. But for $N\geq 2\xi$ these differences
become negligible.
\bibitem[18]{remark4}It will be  discussed in more detail elsewhere.
\end{references}
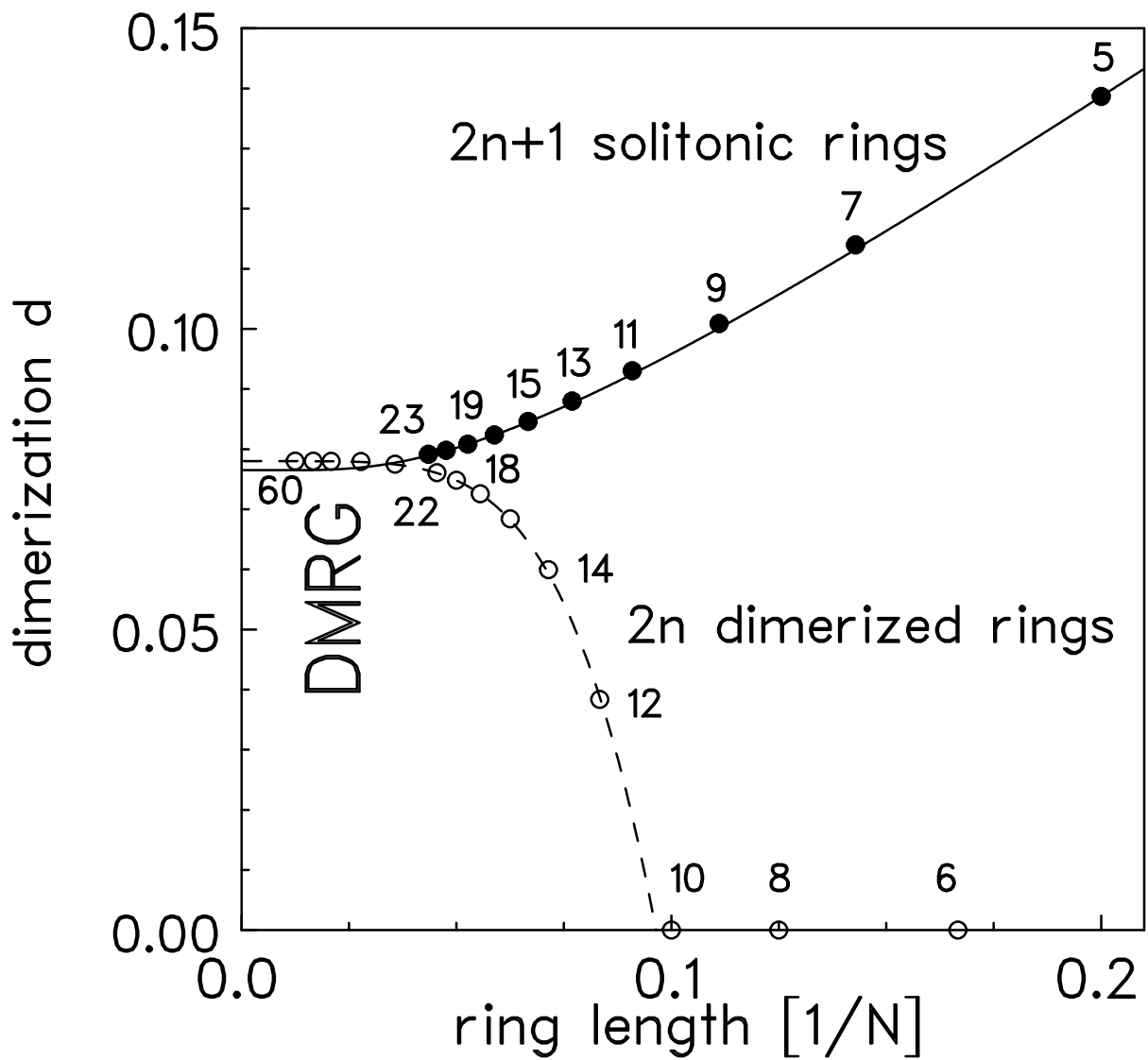
\end{document}